\let\ifarxiv=\iftrue     
\pdfoutput=1
\documentclass[12pt,a4paper]{article}

\ifarxiv\ifnum\pdfoutput=1\else
\PassOptionsToPackage{hypertex}{hyperref}
\PassOptionsToPackage{draft}{graphicx}
\fi\fi

\usepackage{amsmath,amssymb}
\usepackage[bookmarks=true,hyperfigures=true]{hyperref}
\usepackage{graphicx}
\usepackage[nosort]{cite}
\usepackage[bulletsep]{collref}
\usepackage{feynmp}
\usepackage{multirow}
\usepackage{listings}
\usepackage{longtable}
\usepackage{multicol}
\usepackage{hhline}

\usepackage[a4paper,text={173mm,216mm},centering]{geometry}

\allowdisplaybreaks[3]

\numberwithin{equation}{section}

\usepackage[font=small,labelfont=bf,width=0.85\textwidth]{caption}

\makeatletter
\let\old@startsection=\@startsection
\renewcommand{\@startsection}[6]{\old@startsection{#1}{#2}{#3}{#4}{#5}{#6\mathversion{bold}}}
\makeatother

\makeatletter
\newlength{\apb@width}
\newcommand{\autoparbox}[2][c]{\settowidth{\apb@width}{#2}\parbox[#1]{\apb@width}{#2}}

\makeatother

\def \bl  {\begin{align*}}
\def \el  {\end{align*}}
\def \bsp {\begin{equation}\begin{split}}
\def \be  {\begin{equation}}
\def \ee  {\end{equation}}
\def \ba  {\begin{eqnarray}}
\def \ea  {\end{eqnarray}}
\def \baa {\begin{eqnarray*}}
\def \eaa {\end{eqnarray*}}
\def \bb  {\begin {thebibliography} }
\def \eb  {\end{thebibliography}}
\def \lab #1 {\label{#1}}

\newcommand \eqn [1] {(\ref{#1})}

\newcommand{\vectt}[1]{\vec{\tilde{#1}}}

\def \Tr {\mathop{\rm Tr}\nolimits}

\renewcommand{\d}{\delta}

\newcommand{\nn}{\nonumber}

\newcommand{\cL}{{\cal L}}
\newcommand{\cH}{{\cal H}}
\newcommand{\cM}{{\cal M}}
\newcommand{\cC}{{\cal C}}
\newcommand{\unit}{{\mathbf 1}}
\newcommand{\cO}{{\cal O}}
\newcommand{\p}[1]{(\ref{#1})}

\newcommand{\AdS}{$AdS_{5}\times S^{5}$ }
\newcommand{\Nfour}{$\mathcal{N}=4$ }

\def\D{\Delta}

\def\d{\delta}

\def\m{\mu}

\def\n{\nu}

\def\l{\lambda}

\def\p{\partial}

\renewcommand{\title}[1]{\vbox{\center\LARGE{#1}}\vspace{5mm}}
\renewcommand{\author}[1]{\vbox{\center#1}\vspace{5mm}}
\newcommand{\address}[1]{\vbox{\center\em#1}}
\newcommand{\email}[1]{\vbox{\center\tt#1}\vspace{5mm}}

\begin{document}

\begin{titlepage}
\begin{center}
\vspace{5mm}
\hfill {\tt HU-EP-10/85}\\
\vspace{20mm}

\title{\sc On the Spectrum of the  $AdS_{5}\times S^{5}$\\ String at large
  $\lambda$}
\author{\large  Filippo Passerini${}^{1}$, Jan Plef\/ka${}^{1}$, 
Gordon W.~Semenoff${}^{2}$  and Donovan Young${}^{3}$}
\address{
${}^{1}$Institut f\"ur Physik, Humboldt-Universit\"at zu Berlin,\\
Lise-Meitner-Haus, Newtonstra{\ss}e 15, D-12489 Berlin, Germany
\\[0.75cm]
${}^{2}$
Department of Physics and Astronomy, University of British Columbia, \\ 
Vancouver, British Columbia, V6T 1Z1 Canada 
\\[0.75cm]
${}^{3}$
Niels Bohr Institute, Blegdamsvej 17, DK-2100 Copenhagen, Denmark
}
\end{center}

\abstract{We put forth a program for perturbatively quantizing the bosonic sector
of the IIB superstring on $AdS_5\times S^5$ in the large radius of curvature $R$ 
 (i.e. flat-space) limit, and in
light-cone gauge. Using the quantization of the massive particle on $AdS_5\times S^5$ as a
guiding exercise, we read off the correct scaling of the particle coordinates 
 in the large radius limit. 
In the corresponding large $\sqrt{\lambda}=R^{2}/\alpha'$  limit of the string
the oscillator modes must be scaled differently from the zero modes. 
We compute the bosonic string Hamiltonian in this limit which gives the flat-space 
mass-squared operator at leading order, followed by a harmonic oscillator potential in the zero modes at subleading order. Using these ingredients we calculate the leading and sub-leading terms in the conformal dimension of the length-four Konishi state in the large-$\lambda$ limit.
Furthermore we work out the relevant terms of the energy expansion to next-to-next-to leading
order, which are plagued by severe ordering ambiguities. This prevents us from determining
the corrections to the spectrum at order $\lambda^{-1/4}$.}

\vspace{0.75cm} \email{${}^{1}$\{filippo,plefka\}@physik.hu-berlin.de, ${}^{2}$gordonws@phas.ubc.ca,
${}^{3}$dyoung@nbi.dk}

\end{titlepage}

\section{Introduction}

The AdS/CFT correspondence \cite{Maldacena:1998re}\cite{Witten:1998qj,Gubser:1998bc}  
has seen remarkable progress since its
inception over a decade ago. At the forefront of these developments
has been the study of  the spectrum of 
scaling dimensions of local gauge invariant operators 
in planar \Nfour super Yang-Mills theory (SYM) which should be equivalent
to the energy spectrum of quantum type IIB superstrings in an
\AdS background. These advances where made possible by the apparent integrability
\cite{Minahan:2002ve,Beisert:2003yb, Beisert:2003tq, Bena:2003wd}
of this maximally symmetric $AdS_{5}/CFT_{4}$ system 
(for reviews see \cite{Tseytlin:2004cj,Belitsky:2004cz,Beisert:2004ry,Beisert:2004yq,Zarembo:2004hp,Minahan:2006sk,Arutyunov:2009ga} and the very 
recent \cite{Beisert:2010jr}). Generally the 
gauge theory operators or string states fall into representations
of the bosonic subgroup $SO(2,4)\times SO(6)$ of the underlying symmetry
supergroup $PSU(2,2|4)$ which may be labelled by the Cartan charges 
$(E,S_{1},S_{2}; J_{1},J_{2},J_{3})$. Here $E$ is the energy, whereas the $S_{i}$
and $J_{i}$ are angular momenta associated to $AdS_{5}$ and $S^{5}$ respectively.
In limiting situations where subsets of the $S_{i}$ and $J_{i}$ tend to infinity,
corresponding to gauge theory operators of large classical dimension or,
respectively, long spinning strings, 
exact non-perturbative predictions for the spectrum can be made through the
asymptotic Bethe-ansatz (ABA) equations \cite{Beisert:2005fw} in their final form
\cite{Beisert:2006ez}. A particularly striking example
is the large spin $S_{1}\to\infty$ limit from which an exact prediction for the cusp anomalous
dimension of \Nfour SYM could be made and matched to high perturbative orders in the
weak \cite{Bern:2006ew} and strong coupling \cite{Basso:2007wd} expansions.
At strong coupling these asymptotically large Cartan charge 
regimes are accessible through a semiclassical
quantization of the \AdS string in form of a fluctuation expansion around classical spinning string
solutions \cite{Gubser:2002tv,Frolov:2002av}
(for reviews see \cite{Tseytlin:2003ii,Plefka:2005bk}). 
This method has been extensively studied and
refined to various long-string limit scenarios in the literature (see 
e.g.~\cite{Frolov:2006qe,Roiban:2007ju,Beccaria:2008tg,Giombi:2010fa,Giombi:2010zi}).

Recently, substantial progress towards understanding the spectrum of ``short'' gauge theory
operators and short quantum strings was made, i.e.~situations in which the charges 
$J_{i}$ and $S_{i}$ remain 
finite and the asymptotic Bethe equations fail. The prototype gauge theory operator 
in this scenario is the
Konishi operator $\Tr(\phi_{I}\phi_{I})$, where the $\phi_{I}$ denote 
the six real scalars  of \Nfour
SYM. It is the shortest operator with a non-trivial scaling dimension, which has been computed
at weak gauge coupling $\lambda\ll 1$ up to four orders of perturbation theory 
\cite{Fiamberti:2007rj,Fiamberti:2008sh,Velizhanin:2008jd}. 
In fact one usually computes the anomalous scaling dimension
of the Konishi multiplet operator $\Tr[Z,W]^{2}$, with $Z$ and $W$ complex scalars,
differing from the scaling dimensions of the Konishi operator only in its tree-level 
(classical) part. From the integrable systems perspective important parallel progress
was achieved by using two-dimensional field theory inspiration
 to generalize the ABA. On the one hand improvements of the ABA
equations through L\"uscher corrections capturing the wrapping effects at low orders
in $\lambda$ \cite{Ambjorn:2005wa} were efficient enough to extend the 
string/integrability weak coupling
prediction for the Konishi scaling dimensions to the five loop order \cite{Lukowski:2009ce}.
On the other hand the most promising approach appears to be the 
generalization of the ABA description of states on 
a decompactified $R^{1,1}$ string-worldsheet to states on a cylinder
$R\times S^{1}$ by the Thermodynamic
Bethe Ansatz (TBA) and Y-system 
\cite{Arutyunov:2007tc,Arutyunov:2009zu,Gromov:2009tv,Bombardelli:2009ns,Gromov:2009bc,
Arutyunov:2009ur,Arutyunov:2009ux,Gromov:2009zb,Arutyunov:2010gb}.
This approach led to numerical predictions for the scaling dimensions of the Konishi
operator at strong coupling \cite{Gromov:2009zb,Frolov:2010wt}.

It is generally believed
that at large values of the 't Hooft coupling $\lambda$ the energy of the Konishi 
state can be expanded in an asymptotic series in powers of $\lambda^{-1/4}$
\be
\label{formexp}
E_{\text{Konishi}}= c_{-1}\, \lambda^{1/4} + c_{0} + \frac{c_{1}}{\lambda^{1/4}}
+ \frac{c_{2}}{\lambda^{1/2}} + \frac{c_{3}}{\lambda^{3/4}} 
+ \frac{c_{4}}{\lambda}+ \ldots \, . 
\ee
Here the leading term arises from the lowest excited closed string state in flat-space
\cite{Gubser:1998bc} and takes the value $c_{-1}=2$. It has also been shown to arise
from the ABA \cite{Arutyunov:2004vx} in the large $\lambda$ limit. 
In fact such an expansion in $\lambda^{-1/4}$ was shown to arise from the exact solution of the
$\mathfrak{su}(1|1)$-sector truncation of the full \AdS superstring model in \cite{Arutyunov:2005hd},
where in addition the values $c_{-1}=2$ and $c_{0}=0$ were obtained for the first excited states
from both the truncated  $\mathfrak{su}(1|1)$-model and the ABA.
The two independent numerical studies \cite{Gromov:2009zb,Frolov:2010wt} 
based on 
the TBA/Y-system approach have reported
the values\footnote{We present an averaged value of the two reported results.
The value $c_{0}=0$ and the form of the power law expansion
has actually been used as an input for fitting
the numerical data.}
\be
\label{TBAres}
c_{-1}=2.000(2)\, , \qquad 
c_{0}=0 \, , \quad c_{1}= 2.000(3) \, , \qquad c_{2}=-0.0(2) \, , \qquad
c_{3}\sim -{\cal O}(1) \, ,
\ee 
for the length four $\mathfrak{sl}(2)$-sector Konishi descendent $\Tr [D,Z]^{2}$.

Clearly this result poses a challenge for a strong-coupling quantum string analysis.
Here the stumbling block is the full quantization of the IIB
superstring in the $AdS_5\times S^5$ background geometry, a problem
which so far continues to elude solution, and without solution
prevents one from describing string states which are not
semi-classical (i.e.~long) in nature. Nevertheless, semi-classical
string methods have been adapted to this question in the  work of  Roiban and Tseytlin 
\cite{Roiban:2009aa,Tseytlin:2009fw} providing a puzzling conclusion.
They found the coefficients
\be
\label{STres}
c_{-1}=2\, , \qquad 
c_{0}=\Delta_{0}-4 \, , \qquad c_{1}= 1 \, , \qquad c_{2}=0\, .
\ee
Note that for the Konishi multiplet state in question $\Tr[Z,W]^{2}$, the tree-level 
scaling dimensions  is $\Delta_{0}=4$ and hence there is consistency with the numerical
result \eqn{TBAres} for $c_{0}$. A clear
disagreement by a puzzling factor of two, however, resides in the next-to-next-to-leading
order contribution $c_{1}$. 

However, different results than \eqn{STres} were obtained by extrapolating the semi-classical
quantization about spinning folded string solutions in $AdS_{3}$ \cite{Beccaria:2010ry}
\be
\label{STres2}
c_{-1}=2\, , \qquad 
c_{0}=1 \, , \qquad c_{1}=6-8\log 2 \, , \qquad c_{2}=0 \, ,
\ee
with non-integer $c_{1}$. Extrapolating a pulsating string solution to the short
string limit yields yet a different analytical value for $c_{1}$ \cite{Beccaria:2010zn}. Hence
the present status of the strong-coupling string side is certainly unsatisfactory.

What is needed is an efficient perturbative string quantization for the large $\lambda$
limit. In this paper we take the first steps towards this goal
and for the moment consider only the bosonic part of the string
action. We will work in light-cone gauge by using the 
$AdS_{5}$ global time and an azimuthal angle from the $S^{5}$ to build the light-cone
coordinates $x^{\pm}$
\cite{Arutyunov:2005hd,Frolov:2006cc,Arutyunov:2006ak}. 
An alternative approach using a light-cone
gauge entirely within $AdS_{5}$ has been studied in \cite{Metsaev:2000yf,Giombi:2009gd}.

We begin in section \ref{sec:particle} with an exercise: the
quantization of the massive particle\footnote{The quantization of the massless
$AdS_{5}\times S^{5}$ superparticle in AdS light-cone gauge
 was discussed in \cite{Metsaev:1999gz,Horigane:2009qb}.} in the large $MR$ limit of $AdS_5
\times S^5$, with $M$ the mass of the particle and $R$ the curvature radius. 
Here the exact spectrum is known; the recovery of the correct
answer provides us with a proof of concept of our self-consistent perturbative procedure 
and also crucially provides
the correct scaling to apply to the phase space variables in order
achieve the flat-space limit correctly. The scaling we employ is
\be 
\label{scalepar}
\vec x \rightarrow \frac{\vec x}{\sqrt{MR}}, \qquad \vec y
\rightarrow \frac{\vec y}{\sqrt{MR}}, \qquad \vec p \rightarrow
\sqrt{MR}\,\vec p,\qquad \vec q \rightarrow \sqrt{MR}\,\vec q \, ,
\ee
where  $\vec x$ (and, correspondingly $\vec p$) are the
four transverse coordinates (momenta) of $AdS_5$, and $\vec y$ ($\vec
q$) those of $S^5$, see section \ref{sec:particle} for details. 
This form of the scaling may be easily understood by looking at the action of
the massive particle in dimensionless coordinates before any scaling on the
coordinates performed
\be
S_{\text{particle}} = -MR\, \int d\tau \,\sqrt{-\dot x^{\mu}\,  \dot x^{\nu}
\, g_{\mu\nu}(x)}\, .
\ee
The scaling \eqn{scalepar} for the coordinates then yields a pertubatively accessible
theory upon expanding the action in $1/\sqrt{MR}$, as it starts out with a quadratic action in fields independent of the coupling constant $MR$ and all interaction terms then come with
powers of the coupling constant $1/\sqrt{MR}$.

In section \ref{sec:string} we turn to the bosonic sector of the IIB superstring. Here we
apply the same scaling \eqn{scalepar} to the string zero-modes, but a {\it different}
scaling to the internal string oscillations. The scaling of the oscillatory modes
follows the same logic as above: The Nambu-Goto action for the string reads
\be
S_{\text{string}} = -\frac{R^{2}}{\alpha'}\,  \int d\tau \, d \sigma\,
\sqrt{-\det \partial_{r} X^{\mu}\, \partial_{s}X^{\nu}\, g_{\mu\nu}(X)}
\ee
and hence the string coordinates should be scaled by the square-root of the string 
tension $\l^{1/4} \propto
R/\sqrt{\alpha'}$. However, as we are interested in the perturbative spectrum of excited
\emph{massive} string states at large $\lambda$ we should rescale string zero-modes
differently, namely according to  \eqn{scalepar} with $M\sim1/\sqrt{\alpha'}$.
In summary we therefore scale
\bsp
&\vec x \to \frac{1}{\lambda^{1/8}} \vec x,\quad 
\vec p \to \lambda^{1/8} \vec p,\quad
\tilde {\vec X} \to \frac{1}{\lambda^{1/4}} \tilde {\vec X},\quad 
\tilde {\vec P} \to \lambda^{1/4} \tilde{\vec P},\\
&\vec y \to \frac{1}{\lambda^{1/8}} \vec y,\quad 
\vec q \to \lambda^{1/8} \vec q,\quad
\tilde {\vec Y} \to \frac{1}{\lambda^{1/4}} \tilde {\vec Y},\quad 
\tilde {\vec Q} \to \lambda^{1/4} \tilde{\vec Q}\, ,
\end{split}
\ee
where $\tilde {\vec X}$ are the oscillatory modes and $\vec x$ the
zero modes of the $AdS_5$ transverse coordinates, and similarly for
the $S^5$ part of the geometry. We then quantize the string oscillatory
modes in the large $\l$ limit. Note that a priori this perturbative prescription
points at an expansion in powers of $\lambda^{-1/8}$ opposed to
the general expectation \eqn{formexp}. We observe, however, that at least
for the first three non-vanishing orders the expansion turns out to be effectively
in $\lambda^{-1/4}$ as expected.
To leading order one naturally obtains the flat
string spectrum. At subleading order we find a potential for the zero
modes which is a harmonic oscillator in the transverse $AdS_5$
directions and a free particle in the transverse $S^5$ directions, to wit
\be
E^2 - q_\phi^2 = \sqrt{\lambda}\, M^2 +
\lambda^{1/4}\,\left(\vec p^2 + M^2 \vec x^2 + \vec q^2 \right)
+ \lambda^{1/8} H_{1/8} + \l^0 H_0 + \ldots
\ee 
where the operator $E$ is the global $AdS$ energy, $q_\phi$ is the quantum number
for the momentum in the light-cone direction on $S^5$, and $M^2$ is
the string oscillator mass-squared operator in units of $\alpha'$ (i.e. the number operator
for non-zero modes). Indeed by replacing $\sqrt{\l}M^2 \to (MR)^2$,
one obtains exactly the particle spectrum derived in section
\ref{sec:particle} out to order $\l^{1/4}$. We therefore see that to
leading order we obtain a particle whose mass is determined by the
number of string oscillations, which is what one would expect from the
flat-space limit. We have also determined the Hamiltonians at orders
$\l^{1/8}$ and $\l^{0}$, and they are included in the appendices. 
Importantly the Hamiltonian $\lambda^{1/8}H_{1/8}$ is shown to not contribute
to the spectrum of $E^{2}$ down to order $\lambda^{0}$.
We have not, however, been able to fix the normal ordering constants
appearing in these higher-order Hamiltonians $H_{1/8}$ and $H_{0}$, 
and a full treatment
would require the addition of the fermionic degrees of freedom, which
is beyond the scope of the present paper.

It is interesting to use our result to give an answer for the
strong-coupling expansion of the Konishi multiplet's conformal
dimension. It is natural to assume that the string state with two
oscillations in the transverse $S^5$ directions, and two units of
$q_\phi$ should correspond to the length-four member of the Konishi multiplet
\be
{\cal O}(x) = \Tr[Z(x),W(x)]^{2},
\ee 
where $Z$ and $W$ are two of the complex scalar fields of ${\cal N}=4$
supersymmetric Yang-Mills theory. We would therefore have that the
eigenvalue\footnote{In fact, the eigenvalue of $M^2$ should be 4
only once the fermionic modes have been added. These remove the normal
ordering constants from the bosonic modes which would otherwise give
$M^2=0$ for the level-two states.} of $M^2$ is 4, coming from two 
transverse $S^{5}$ string oscillator excitations, while that
of $q_\phi$ is 2. This gives
\be 
E = 2 \l^{1/4} + \frac{1}{4} \left(\vec p^2 + 4\, \vec x^2 + \vec
q^2 \right)\l^0 + \ldots 
\ee 
The natural ground state for the zero modes is a harmonic oscillator
ground state for the $AdS_5$ transverse directions tensored with a
plane wave of zero momentum in the $S^5$ transverse directions. If we
had included the fermionic degrees of freedom, we expect that the
harmonic oscillator would be generalized to the superharmonic
oscillator, which would give a vanishing ground state energy. This
is consistent with the results of \cite{Gromov:2009zb,Frolov:2010wt}
of eqn.~\eqn{TBAres}, 
where no $\l^0$ term is found in the expansion
for $E$. Hence if we introduce the gauge theory operator- string theory state 
correspondence
\be
\label{osm}
\mathcal{O}(x) \qquad \Leftrightarrow \qquad
\beta_{-1}^{W}\, {\tilde \beta}_{-1}^{W}\, |0\rangle
\ee
with $\beta_{n}^{W}:= \beta_{n}^{1}+ i \beta_{n}^{2}$ in the notation
of section 3 and where $|0\rangle$ denotes the oscillator and zero-mode groundstate, we
find agreement with \eqn{TBAres} and \eqn{STres} to order $\lambda^{0}$.
Note also that in \eqn{STres2} as well as in the further `short string'
examples quoted in \cite{Beccaria:2010zn} there is a non-zero integer for $c_{0}$
which points at a zero-mode sector excitation in our picture.

It would be very interesting to continue the calculation to
higher orders. In appendix \ref{app:18} we show that for the states of
interest in \eqn{osm} the $H_{1/8}$ term does not contribute. We provide the $H_0$
Hamiltonian in appendix \ref{app:0}. Supplemented   by the fermionic
degrees of freedom, this term is sufficient to calculate the ${\cal
O}(\l^{-1/4})$ contribution to the energy $E$, which would provide a
verdict on the differing values of $c_{1}$ from the TBA/Y-system 
\eqn{TBAres} \cite{Gromov:2009zb,Frolov:2010wt} versus semi-classical strings 
\eqn{STres} \cite{Roiban:2009aa} and \eqn{STres2}
\cite{Beccaria:2010ry}. We leave this calculation to further
research.

\section{Massive particle on  $AdS_5\times S^5$}
\label{sec:particle}

Using a generalization of the methods in \cite{Dorn:2005ja,Dorn:2010wt}, the
quantum spectrum of the massive particle on $AdS_5\times S^5$ may be
precisely obtained \cite{privateGeorge}
\be
E_{J,n} = 2 + n + \sqrt{4 + J(J+4) + (MR)^2}, 
\ee
where $M$ is the mass of the particle and $R$ the common radius of
$AdS_5\times S^5$, $J$ is the total angular momentum on $S^5$, and $n
\geq0$ is a level number corresponding to excitations of the
wavefunction in the $AdS_5$ directions. Taking the large $MR$ limit,
one obtains
\be\label{exact}
E_{J,n} = (MR) + n + 2 +\frac{1}{2MR}\Bigr(J(J+4)+4\Bigl)+
{\cal O}((MR)^{-2}).
\ee

As a warm-up to the string case, we would like to reproduce this
result via a direct quantization of the particle action, using
coordinates compatible with the standard light-cone coordinates we
will use in the next section. We therefore use the $AdS_5$ embedding
coordinates $X^{A}$ with metric $\eta_{AB}=(-1,-1,1,1,1,1)$ and index
$A=(0',0,i)$.  We introduce polar coordinates for the $0-0'$
directions $X_{0}=r\cos\theta$ and $X_{0'}=r\sin\theta$ and we denote
the remaining coordinates as $X^{i}=x^i$.  For the $S^{5}$ space, the
embedding coordinates are $Y^{\hat{A}}$ with metric
$\eta_{\hat{A}\hat{B}}=(1,1,1,1,1,1)$ and index $\hat{A}=(0',0,i)$.
Also for this space, we introduce polar coordinates for the $0-0'$
directions $Y_{0}=s\cos\phi$ and $Y_{0'}=s\sin\phi$ and we denote the
remaining coordinates as $Y^{i}=y^i$.  The phase space form of the
massive $AdS_{5}\times S^{5}$ particle Lagrangian then becomes
\begin{align}
  \cL &= (\dot r\, p_{r} +\dot \theta\, p_{\theta} +
  \dot{\vec{x}}\cdot \vec p+\dot s\, q_{s} +\dot \phi\, q_{\phi} +
  \dot{\vec y}\cdot q )
  -\frac{e}{2} ( -p_{r}^{2}-\frac{p_{\theta}^{2}}{r^{2}} + \vec
  p^{2}+q_{s}^{2}+\frac{q_{\phi}^{2}}{s^{2}} + \vec q^{2}+ M^{2})
  \nn\\
  & \quad +\frac{\lambda_{1}}{2}\, (\vec x^{2} -r^{2}+R^{2})
  +\lambda_{2}\, (r\,p_{r}+\vec x\cdot \vec p\, )\,
  +\frac{\lambda_{3}}{2}\, (\vec y^{2} +s^{2}-R^{2}) +\lambda_{4}\,
  (s\,q_{s}+\vec y\cdot \vec q\, )\,
\end{align}
where the terms proportional to $\lambda_{2}$ and $\lambda_{4}$ are
secondary constraints arising from the primary constraints
$X^{A}\,X_{A}+R^{2}=0$ and $Y^{\hat A}\,Y_{\hat A}-R^{2}=0$. We can
solve the Hamiltonian constraint for $p_{\theta}$, which is the
space-time energy whose spectrum we are interested in \be
p_{\theta}^{2}=r^2\, [-p_{r}^{2} + \vec
p^{2}+q_{s}^{2}+\frac{q_{\phi}^{2}}{s^{2}} + \vec q^{2}+ M^{2}] \,.
\ee The primary and secondary constraints can be solved for the
variables $r$, $p_{r}$, $s$ and $q_{s}$ \be r=\sqrt{R^2+\vec x^{2}}\,
, \qquad p_{r}=-\frac{\vec x\cdot \vec p}{\sqrt{R^{2}+\vec
    x^{2}}}\,\qquad s=\sqrt{R^2-\vec y^{2}}\, , \qquad
q_{s}=-\frac{\vec y\cdot \vec q}{\sqrt{R^{2}-\vec y^{2}}}\, .  \ee 

In the following it will be useful to work in dimensionless quantities
for the coordinates and momenta: \be \tilde{x}_{i}=\frac{x_{i}}{R}\, ,
\qquad \tilde{p_{i}}=R\, p_{i} \qquad \tilde{y}_{i}=\frac{y_{i}}{R}\,
, \qquad \tilde{q_{i}}=R\, q_{i} \,, \ee in terms of which the
dimensionless energy squared takes the form \be p_{\theta}^{2}
=(1+\tilde{\vec x}^2)\left[\tilde{\vec p}^2+\tilde{\vec q}^2+(MR)^{2}
- \frac{(\tilde{\vec x}\cdot\tilde{\vec p})^2}{1+\tilde{\vec x}^2}
+\frac{(\tilde{\vec y}\cdot\tilde{\vec q})^2}{1-\tilde{\vec y}^2}
+\frac{q_{\phi}^2}{1-\tilde{\vec y}^2} \right]\, .
\label{en1}
\ee We note the Dirac-brackets originating from the $\lambda_{1}$,
$\lambda_{2}$, $\lambda_{3}$ and $\lambda_{4}$ constraints:
\begin{eqnarray}
\{\theta, p_{\theta}  \} = 1\, , \qquad
\{ \tilde{x}_{i}, \tilde{p}_{j}  \}_{D}  = \delta_{ij} +\tilde{x}_{i}\,\tilde{x}_{j}\, ,
\qquad
\{ \tilde{p}_{i}, \tilde{p}_{j}  \}_{D}  = \tilde{x}_{i}\, \tilde{p}_{j} -
\tilde{x}_{j}\, \tilde{p}_{i}\, ,\\  \{\phi, p_{\phi}  \} = 1\, , \qquad
\{ \tilde{y}_{i}, \tilde{q}_{j}  \}_{D}  = \delta_{ij} -\tilde{y}_{i}\,\tilde{y}_{j}\, ,
\qquad
\{ \tilde{q}_{i}, \tilde{q}_{j}  \}_{D}  = -\tilde{y}_{i}\, \tilde{q}_{j} +
\tilde{y}_{j}\, \tilde{q}_{i} \, .
\end{eqnarray}
These can be nicely mapped to a canonical system, which we shall in a
slight abuse of notation again denote by $(x_{i},y_{i},p_{i},q_{i})$
via \be \tilde{x}_{i}=x_{i}\, ,\qquad \tilde{p}_{i}=p_{i}+x_{i}\,
(\vec x\cdot \vec p)\,\qquad \tilde{y}_{i}=y_{i}\, ,\qquad
\tilde{q}_{i}=q_{i}-y_{i}\, (\vec y\cdot \vec q) ,
\label{tildedefs}
\ee with canonical brackets \be\label{pois} \{x_{i},p_{j}\} =\delta_{ij}\,\qquad
\{y_{i},q_{j}\} =\delta_{ij}.  \ee Inserting the representation
\eqn{tildedefs} into \eqn{en1} yields the space-time energy
\begin{eqnarray}
  p_{\theta}^{2} =(1+\vec x^{2})\left( (MR)^{2} + \vec{p}^{2} + (\vec{p}\cdot \vec{x})^{2} + \vec{q}^{2} -(\vec{q}\cdot \vec{y})^{2} +\frac{q_{\phi}^2}{1-\vec y^{2}} \right)\, .
\end{eqnarray}
The  $S^5$ factor of the  wavefunction  $\Psi_J$,  in   the  coordinate  system  defined  in \eqn{tildedefs} satisfies the following    equation
\begin{eqnarray}
 \left( \vec{q}^{2} -(\vec{q}\cdot \vec{y})^{2} +\frac{q_{\phi}^2}{1-\vec y^{2}} \right)\Psi_{J}=J(J+4)\Psi_{J}
\end{eqnarray}
and this implies that  the space-time energy can be written as 
\begin{eqnarray}\label{jen}
  p_{\theta}^{2} =(1+\vec x^{2})\left( (MR)^{2} + \vec{p}^{2} + (\vec{p}\cdot \vec{x})^{2} + J(J+4) \right)\, .
\end{eqnarray}
Hence we see that the energy $p_{\theta}$ is bounded from below by $MR$.

The equation \eqn{jen} shows  that, as expected,  the  contribution of the  $S^5$ excitations is implemented by a shift of the mass-squared: $M^2 \to M^2 + J(J+4)$. 

To study the particle in the large radius
approximation it is convenient to rescale the  $AdS$ variables in the
following way \be \vec x \rightarrow \frac{\vec x}{\sqrt{MR}} \qquad \qquad \vec p \rightarrow
\sqrt{MR}\,\vec p\ee so
that the squared energy can be expanded as

\bsp \label{enclassical}
p_{\theta}^{2} =(MR)^{2} +(MR)\big( \vec{p}^{2} + \vec{x}^{2}\big)
+\vec{p}^{2}\, \vec{x}^{2} + (\vec{p}\cdot \vec{x})^{2}+J(J+4)+
{\cal O}[(MR)^{-1}]
\end{split}
\ee
This gives
\be\label{ene}
E = p_\theta = MR + \frac{1}{2}\big( \vec{p}^{2} + \vec{x}^{2}\big) + {\cal O}[(MR)^{-1}]
\ee
In section \ref{sec:string} we will recover the ${\cal O}[(MR)^0]$
result above for the zero-modes of the bosonic string. Now, promoting the  (\ref{pois}) to their quantum analogues, for the $AdS$ coordinates we have
\be
[\hat x_i,\hat p_j] = i\d_{ij},\quad \hat x_i =
\frac{1}{\sqrt{2}}(a_i+a_i^\dag),\quad
\hat p_i = -\frac{i}{\sqrt{2}}(a_i - a_i^\dag),
\ee
and therefore
\bsp
E_n &= (MR) 
+ a_i^\dag a_i + \d_{ii}/2 + \ldots \\
& = (MR) 
+ \hat n + 2 + \ldots.
\end{split}
\ee
which matches (\ref{exact}) at the leading two orders. We can go
further and tackle the subleading terms in
(\ref{enclassical}). Introducing constants $d_{i}$ to capture ordering
ambiguities originating
from $x$-$p$ self-contractions, we have
\begin{align}
({p}_{i})^{2}\, ({x}_{i})^{2} + ({p}_{i}\, {x}_{i})^{2} &\to
({\hat x}_{i})^{2}\, ({\hat p}_{i})^{2} + ({\hat x}_{i}\, {\hat p}_{i})^{2} + 
i\, d_{1}\,  ({\hat x}_{i}\, {\hat p}_{i})
+d_{2}\, \unit\, . 
\end{align}
Imposing the hermiticity of the operator $(\hat p_{\theta})^{2}=(\hat p_{\theta}^{\dagger})^{2}$
fixes $d_{1}=-6$. One finds
\bsp
{\hat p}_{\theta}^{2} &= (MR)^2 + MR\, \big (\, 2\, \hat n +4 \, \big ) \\
&\qquad + \big (\, 12+d_{2}+ 4\, \hat n +\hat n^{2}
-\frac{1}{2}\, [\, (a_{i}^{\dagger}\, a_{i}^{\dagger})^{2}+(a_{i}\, a_{i})^{2}\, ]+J(J+4) \big )
+ \cO [(MR)^{-1}] 
\end{split}
\ee
where $\hat n := a^{\dagger}_{i}\, a_{i}$. We first note that the non-diagonal term at order
$\cO[1]$ can be removed by a unitary transformation up to terms of order $\cO[(MR)^{-1}]$:
\bsp
{\hat p}_{\theta}^{2} \to e^{\hat V/(MR)}\, {\hat p}_{\theta}^{2}\, e^{-\hat V/(MR)} 
&= (MR)^2 + MR\, \big (\, 2\, \hat n +4 \, \big ) \\
&\qquad + \big (\, 12+d_{2}+ 4\, \hat n +\hat n^{2} +J(J+4) \big ) + \cO [(MR)^{-1}]
\, ,
\end{split}
\ee
with the anti-hermitian operator
\be
\hat V= -\frac{1}{16}\, \Bigl [ \, (a_{i}^{\dagger}\, a_{i}^{\dagger})^{2}-(a_{i}\, a_{i})^{2} \, \Bigr ]\, .
\ee
Thus we find the spectrum $E_{n}$ of $p_{\theta}$:
\begin{align}
E_{n}^{2} &= (MR)^{2}+ MR\, (2n+4) + 12+d_{2}+4n+n^{2} +J(J+4)+ \cO[(MR)^{-1}], \nn\\
E_{n} &=  MR+ n+2 + \frac{8+d_{2}+J(J+4)}{2 MR} + \cO[(MR)^{-2}]\, 
 \, ,\nn
\end{align}
which  agrees with (\ref{exact})   for the choice $d_{2}=-4$ of the normal ordering
constant. Indeed this value of the ordering constant $c_{2}$ can be shown to be unambiguously
determined by the closure of the $SO(2,4)$ quantum symmetry algebra of the 
$AdS_{5}\times S^{5}$ particle problem.

\section{Bosonic string on  $AdS_5\times S^5$}
\label{sec:string}

In the previous section  we saw how a rescaling of the transverse particle
coordinates and momenta
\be
(x_i,\, y_i) \to \frac{1}{\sqrt{MR}} (x_i,\, y_i), 
\qquad (p_i,\, q_i) \to \sqrt{MR}\, (p_i,\,q_i)
\ee
led to a recovering of the quantum spectrum in the $MR \to \infty$
limit.   In
generalizing to the string we keep this scaling for the string
zero-modes, while scaling the oscillating string modes with  $\lambda^{1/4} \sim
R/\sqrt{\alpha'}$. As described in the  introduction, this is the  scaling  that produces  a pertubatively accessible
theory.   What we will find is the flat-space string spectrum at leading order,
and the particle spectrum found above at the first subleading
order, where the mass of the particle is given by the flat space mass of the string.

Let us consider the coordinate system where the $AdS_5\times S^5$ metric
is 
\begin{equation}
ds^2=-\left( \frac{1+z^2/4}{1-z^2/4}\right)^2 dt^2+ 
\frac{d\vec z\cdot d\vec z}{(1-z^2/4)^2}
+\left( \frac{1-y^2/4}{1+y^2/4}\right)^2d\phi^2 + 
\frac{d\vec y\cdot d\vec y}{(1+y^2/4)^2}.
\end{equation}
This can be obtained by a simple coordinate transformation of the
coordinates in section \ref{sec:particle}
\begin{equation}
\vec x \to \frac{\vec z}{1-z^2/4}
~~,~~
\vec {y} \to \frac{\vec y}{1+y^2/4}.
\end{equation}
We will take the convention where the action is 
\begin{equation}
 S=\int d\tau \int_0^{2\pi}\frac{d\sigma}{2\pi}\left[P_T\dot T+P\cdot\dot Z+Q_\phi\dot\phi+Q\cdot\dot Y
-\eta_1{\cal S}-\eta_2{\cal T}\right].
\end{equation} 
The constraints which are enforced by the Lagrange multipliers $\eta_i$ are
\begin{equation}
{\cal S}=0 ~~,~~ 
{\cal T}=0,
\end{equation}
where
\begin{eqnarray}
&{\cal S}=P_T T'+Q_\phi \phi' +P\cdot Z'+\vec Q\cdot\vec Y',\\
&{\cal T}=\frac{1}{\sqrt{\lambda}}\left[
-\left( \frac{1-\vec Z^2/4}{1+\vec Z^2/4}\right)^2P_T^2 
+(1-\vec Z^2/4)^2\vec P^2 +\left( \frac{1+\vec Y^2/4}{1-\vec Y^2/4}\right)^2Q_\phi^2
+(1+\vec Y^2/4)^2\vec Q^2 \right]\nonumber \\
&+\sqrt{\lambda}
\left[-\left( \frac{1+\vec Z^2/4}{1-\vec Z^2/4}\right)^2 
{T'}^2+ 
\frac{(\vec Z')^2}{(1-\vec Z^2/4)^2}
+\left( \frac{1-\vec Y^2/4}{1+\vec Y^2/4}\right)^2(\phi')^2 + 
\frac{(\vec Y')^2}{(1+\vec Y^2/4)^2}\right].
\end{eqnarray}
We now introduce light-cone coordinates using global time $T$ and the azimuthal
angle $\phi$
\begin{eqnarray}
X_-=\phi- T &,& X_+=\frac{1}{2}(T+\phi). \label{lc1}\\
P_-=Q_\phi + P_T &,& P_+=\frac{1}{2} (Q_\phi -P_T). \label{lc2}\\
T = X_+ - \frac{1}{2} X_- &,& \phi=X_+ +\frac{1}{2}X_-. \label{lc3}\\
P_T = \frac{1}{2}P_- - P_+ &,& Q_\phi =\frac{1}{2}P_- + P_+. 
\label{lc4}
\end{eqnarray}
Then,
we impose the light-cone gauge conditions,
\begin{equation}
P_+=p_+
~~,~~
X_+=x_+ + p_+\tau,
\end{equation}
where $x_+$ and $p_+$ are $(\tau,\sigma)$-independent.
With these conditions, 
the constraints are
\begin{align}
0=&p_+X_-' +P\cdot Z'+\vec Q\cdot\vec Y'~\to~
~~ X_-' = - \frac{1}{p_+}\left(P\cdot Z'+\vec Q\cdot\vec Y'\right), \label{virasoro1}
\\
0=&
\left[
\left( \frac{1+\vec Y^2/4}{1-\vec Y^2/4}\right)^2
-\left( \frac{1-\vec Z^2/4}{1+\vec Z^2/4}\right)^2\right] 
\left(\frac{P_-^2}{4}+p_+^2\right)
+\left[\left( \frac{1+\vec Y^2/4}{1-\vec Y^2/4}\right)^2
+\left( \frac{1-\vec Z^2/4}{1+\vec Z^2/4}\right)^2\right]
P_-p_+ 
\nonumber \\
&+(1-Z^2/4)^2\vec P^2 +(1+\vec Y^2/4)^2 \vec Q^2
+\lambda\left[\frac{(\vec Z')^2}{(1-\vec Z^2/4)^2}
+\frac{(\vec Y')^2}{(1+\vec Y^2/4)^2}\right]
\nonumber \\ 
&+\lambda\left[\left( \frac{1-\vec Y^2/4}{1+\vec Y^2/4}\right)^2
-\left( \frac{1+\vec Z^2/4}{1-\vec Z^2/4}\right)^2\right]
\frac{1}{4p_+^2}\left(\vec P\cdot\vec Z'+\vec Q\cdot \vec Y'\right)^2 ,
\label{virasoro2}
\end{align}
where, for convenience, the right-hand-side of (\ref{virasoro2}) is
${\cal T}$ rescaled by a factor of ${\sqrt{\lambda}}$.
Here, we have solved the constraint ${\cal S}=0$ as
indicated and plugged the solution into the second constraint to obtain
(\ref{virasoro2}).  In the following, we shall use lower case letters to denote the worldsheet averages of coordinates and momenta
$x_\mu=\int_0^{2\pi}\tfrac{d\sigma}{2\pi}X_\mu(\sigma )$  and  $p_\mu=\int_0^{2\pi}\tfrac{d\sigma}{2\pi}P_\mu(\sigma )$.

We must now solve (\ref{virasoro2}) for the
remaining variable $P_-(\sigma)$.  This will be done perturbatively
about the large $\lambda$ limit, according to a scheme which we
outlined in the introduction.  Together with finding $P_-(\sigma)$, we will find
expressions for the momenta $p_+$ and $p_-$.

We begin by recalling that $p_+$ and $p_-$ are not independent, they are related by the second equation in (\ref{lc4}),
\begin{equation}q_\phi=\tfrac{1}{2}p_- + p_+ \, .
\end{equation}  
Here, $q_\phi$ is conjugate to the
zero mode of the angle coordinate $\phi$ and  its spectrum is integers.
We will be interested in states where the magnitude of these integers
is of order $\lambda^0$.   
We will see
that, as a consequence, the leading terms in $p_+$ and $p_-$ must be of order $\lambda^{1/4}$, and
the asymptotic expansion of these quantities is generically in powers of $\lambda^{-1/8}$.
From $p_+$ and $p_-$, we will deduce the spectrum of the squared string energy, $p_T^2$ where
\begin{equation}\label{pt}
p_T=\tfrac{1}{2}p_--p_+~~,~~p_T^2=q_\phi^2 - 2p_+p_-,
\end{equation}
where we have used Eq.~(\ref{lc2}).

To proceed, we will scale the fields as 
\begin{eqnarray}
\vec Z(\sigma,\tau)=\frac{1}{\lambda^{\frac{1}{8}}}\vec z(\tau)
+\frac{1}{\lambda^{\frac{1}{4}}}\vec{\tilde Z}(\sigma,\tau) &,&
\vec P(\sigma,\tau)=\lambda^{\frac{1}{8}}\vec p(\tau)
+\lambda^{\frac{1}{4}}\vec{\tilde P}(\sigma,\tau)\, , \\
\vec Y(\sigma,\tau)=\frac{1}{\lambda^{\frac{1}{8}}}\vec y(\tau)
+\frac{1}{\lambda^{\frac{1}{4}}}\vec{\tilde Y}(\sigma,\tau) &,&
\vec Q(\sigma,\tau)=\lambda^{\frac{1}{8}}\vec q(\tau)
+\lambda^{\frac{1}{4}}\vec{\tilde Q}(\sigma,\tau).
\end{eqnarray}
Here, we have separated the zero modes,  $(\vec p,\vec z,
\vec q,\vec y)$ from the internal oscillations of the string which we denote as $(\vec{\tilde P},
\vec{\tilde Z},\vec{\tilde Q},\vec{\tilde Y}$) and  which are constrained by
$$
\int d\sigma \vec{\tilde P} = 
\int d\sigma \vec{\tilde Z} = \int d\sigma \vec{\tilde Q}=\int d\sigma \vec{\tilde Y}=0.
$$
The scaling of zero modes is consistent with that in Section 2.  A priori this scaling suggest an expansion of the energy in powers of $\lambda^{-1/8}$.

We shall find that $P_-$ and $p_+$ scale as $\lambda^{\frac{1}{4}}$ for
large $\lambda$.  Then, to the leading order in large $\lambda$, (\ref{virasoro2}) becomes
\begin{eqnarray}\label{leadingorder}
-2P_-p_+ =\lambda^{\frac{1}{2}}{\cal M}^2(\sigma)+\ldots,
\end{eqnarray}
where the dots indicate terms of order less than $\lambda^{\frac{1}{2}}$ and the
mass operator-density is 
\begin{equation}
{\cal M}^2(\sigma)=\left[
\vec{\tilde P}^2 +\vec {\tilde Q}^2
+(\vec{\tilde Z}')^2
+(\vec{\tilde Y}')^2\right].
\end{equation}
Then, remembering that $2p_+p_-=q_\phi^2-p_T^2$, we get, after integrating (\ref{leadingorder}) 
over $\sigma$, 
\begin{equation}
p_T^2=q_\phi^2+\lambda^{\frac{1}{2}}M^2+{\cal O}(\lambda^{\frac{1}{4}}),
\end{equation}
where
\begin{equation}
M^2\equiv\frac{1}{2\pi}\int_0^{2\pi}d\sigma{\cal M}^2(\sigma)
=\frac{1}{2\pi}\int_0^{2\pi}d\sigma\left[
\vec{\tilde P}^2 +\vec {\tilde Q}^2
+(\vec{\tilde Z}')^2
+(\vec{\tilde Y}')^2\right],
\end{equation}
is the flat-space mass operator.
For the states of interest to us,  $q_\phi$ will be of order one.  
Also, we learn that 
\begin{eqnarray}
p_+&=&p_+^{(0)} \lambda^{1/4} + p_+^{(2)},
\\
P_-(\sigma)&=&-\lambda^{\frac{1}{4}}\frac{{\cal M}^2(\sigma)}{2p_+^{(0)}}+\ldots,
\\
p_-&=&-\lambda^{\frac{1}{4}}\frac{M^2}{2p_+^{(0)}}+\ldots.
\end{eqnarray}
where the three dots in each of the above formulae denote
corrections of order at least $\lambda^{\frac{1}{8}}$.
We leave the constants $p_{+}^{(0)}$ and $p_{+}^{(2)}$ of the $p_{+}$ expansion undetermined
for the moment\footnote{Note that a term of the form $\lambda^{\frac{1}{8}}\,
p_+^{(1)}$ does not appear.  This can be seen using equation
(\ref{lc2}) for $p_\pm$ and recalling that we are interested in states
where $q_\phi$ is of order one.}. They will be shown to follow at each order
in the $\lambda^{-1/8}$ expansion from a 
self-consistency analysis.

Now, we must solve the equation for $P_-$ to the next order. 
 There is an order $\lambda^{\frac{1}{8}}$ contribution which is arises by expanding 
to the next order. Because of the orthogonality conditions, 
$\int d\sigma \tilde P^i=0$, $\int d\sigma \tilde Q^j=0$,
the result does not contribute to $p_-$ or $p_+$, but
it must be taken into account in $P_-(\sigma)$.  In summary, so far, we have
\begin{eqnarray}
P_-(\sigma)&=&-\lambda^{\frac{1}{4}}\frac{{\cal M}^2(\sigma)}{2p_+^{(0)}}
-\lambda^{\frac{1}{8}}\left [ \frac{1}{p_+^{(0)}}\left( \vec p\cdot\vec{\tilde P}+\vec q\cdot\vec
{\tilde Q}\right)
\right ]+\ldots,  \\
p_-&=&-\lambda^{\frac{1}{4}}\frac{M^2}{2p_+^{(0)}}+\ldots,
\end{eqnarray}
where the dots stand for terms of order $\lambda^0$ and
higher. Now, we are ready to solve the next
non-trivial order.  For this, we expand (\ref{virasoro2}) as
\begin{align}
- 2 P_- p_+ =&
\lambda^{\frac{1}{2}}  {\cal M}^2 +
2\lambda^{\frac{3}{8}}
\left[
\vec p\cdot\vec{\tilde P}+\vec q\cdot\vec{\tilde Q}
\right]
+\lambda^{\frac{1}{4}}{\cal H}_{1/4}(\sigma) +\ldots
\label{virasoro4}
\end{align}
where we have introduced
\begin{align}\nonumber
 {\cal H}_{1/4}(\sigma)=& 
\vec p^2+\vec q^2 +  
\left( \vec z^2+\vec y^2\right) (p_+^{(0)})^2
-\frac{{\vec y}^2}{2}  \left( {\cal M}_\alpha^2 +2{\vectt{Y'}}^2
\right)\\&+\frac{{\vec z}^2}{2} \left( {\cal M}_\beta^2  +2{\vectt{Z'}}^2
\right)
+\frac{{\vec y}^2 + {\vec z}^2}{16 (p_+^{(0)})^2}
\left( {\cal M}^4 -4 {\cal C}^2\right),
\label{pminusdensitytosecondorder}
\end{align}
with
\begin{align}
\cM_{\alpha}^2&\equiv \vectt{P}^{2}+{\vectt{Z}'}^{2}\, , \quad
\cM_{\beta}^2\equiv \vectt{Q}^{2}+{\vectt{Y}'}^{2}\, , \quad
\cM^2=\cM_{\alpha}^2+\cM_{\beta}^2\, ,\quad
\cC \equiv \vectt{P}\cdot\vectt{Z}' + \vectt{Q}\cdot\vectt{Y}'\,.
\end{align}
From this equation we learn that 
\begin{align}
\label{pminusdensitytosecondorder}
P_-(\sigma)=& -\lambda^{\frac{1}{4}}\frac{{\cal M}^2}{2p_+^{(0)}} - \lambda^{\frac{1}{8}}
\frac{1}{p_+^{(0)}}\left(\vec p\cdot \vec{\tilde P} + \vec q\cdot\vec{\tilde Q}\right)
+p_+^{(2)}\frac{ {\cal M}^2(\sigma)}{2(p_+^{(0)})^2} 
\nonumber \\
&-\frac{1}{2p_+^{(0)}}
 {\cal H}_{1/4}(\sigma)+\ldots,
\\
p_-=&-\lambda^{\frac{1}{4}}\frac{M^2}{2p_+^{(0)}}
+p_+^{(2)}\frac{ M^2}{2(p_+^{(0)})^2} - \frac{1}{2p_+^{(0)}} H_{1/4} 
+\ldots ,
\end{align}
where 
\begin{eqnarray}
H_{1/4}&=&\int_0^{2\pi}\frac{d\sigma}{2\pi}
 {\cal H}_{1/4}(\sigma)\, .
\label{hquarter}
\end{eqnarray}
The three dots at the end of (\ref{pminusdensitytosecondorder}) represent terms of order at least
$\lambda^{-\tfrac{1}{8}}$. The energy squared $p_{T}^{2}$ has a simpler expression 
and reads
\be
\label{sf2}
p_{T}^{2} = \lambda^{1/2}\, M^{2} + \lambda^{1/4}\, H_{1/4} + \lambda^{1/8}\, H_{1/8}
+ \lambda^{0}\, (H_{0} + q_{\phi}^{2}) + \cO(\lambda^{-1/8})
\ee

The procedure which we are following here can be iterated to a systematic computation of the
classical $p_+$, $p_-$ and $P_-(\sigma)$ to any order.  In the Appendix we work out the operators that
are needed to compute $p_T^2$ to orders $\lambda^{1/8}$ and $\lambda^{0}$, ${ H}_{1/8}$
and ${H}_0$, respectively. 

\subsection{Quantization and the string spectrum}

The string coordinates and momenta obey the equal-time Poisson brackets
\begin{equation}
 \left\{ Z^i(\sigma,\tau),P^j(\sigma',\tau)\right\}=2\pi\delta(\sigma-\sigma') \delta^{ij},~
 \left\{ Y^{'i}(\sigma,\tau),Q^{j'}(\sigma',\tau)\right\}=2\pi\delta(\sigma-\sigma')\delta^{i'j'}.
\end{equation}
We solve these and diagonalize the flat space mass operator $M^2$ with the oscillator expansion
\begin{eqnarray}
 \tilde
 Z^i(\sigma,\tau)&=&\frac{i}{\sqrt{2}}\sum_{n\neq0}\left[\frac{\alpha^i_n(\tau)}{n}e^{-in \sigma}+
\frac{\tilde\alpha^i_n(\tau)}{n}e^{in\sigma}\right], \\
\tilde P^i(\sigma,\tau)&=&\frac{1}{\sqrt{2}}\sum_{n\neq0}\left[\alpha^i_n(\tau) e^{-in\sigma}+
\tilde\alpha^i_n(\tau)e^{in\sigma}\right],  \\
 \tilde Y^{i'}(\sigma,\tau)&=&\frac{i}{\sqrt{2}}\sum_{n\neq0}\left[\frac{\beta_n^{i'}(\tau)}{n}
 e^{-in\sigma}+
\frac{\tilde\beta^{i'}_n(\tau)}{n}e^{in\sigma}\right], \\
\tilde Q^{i'}(\sigma,\tau)&=&\frac{1}{\sqrt{2}}\sum_{n\neq0}\left[\beta^{i'}_n(\tau)e^{-in\sigma}+
\tilde\beta^{i'}_n(\tau)e^{in\sigma}\right],
\end{eqnarray}
where the non-vanishing equal-time oscillator brackets are
\bsp
&\left\{z^i(\tau),p^j(\tau)\right\}=\delta^{ij},~~
 \left\{\alpha_m^i,\alpha_n^j\right\}=-im\delta_{m+n}\delta^{ij},~~
 \left\{\tilde\alpha^i_m,\tilde\alpha^j_n\right\}=-im\delta_{m+n}\delta^{ij}, \\
&\left\{y^{i'}(\tau),q^{j'}(\tau)\right\}=\delta^{i'j'},~~
 \left\{\beta_m^{i'},\beta_n^{j'}\right\}=-im\delta_{m+n}\delta^{i'j'},~~
 \left\{\tilde\beta^{i'}_m,\tilde\beta^{j'}_n\right\}=-im\delta_{m+n}\delta^{i'j'}.
\end{split}
\ee
The Virasoro generators are defined in such a way
to exclude any zero modes, i.e.
\begin{equation}
L_n \equiv \frac{1}{2} \sum_{\substack{m=-\infty\\m\neq n,0}}^\infty
\left(\vec \alpha_{n-m} \cdot \vec \alpha_m + \vec \beta_{n-m} \cdot
  \vec \beta_m \right) .
\end{equation}
With this convention,
\begin{equation}
  M^2=\sum_{n\neq0} \left[
    \vec\alpha_{-n}\cdot\vec\alpha_n+\vec{\tilde\alpha}_{-n}\cdot\vec{\tilde\alpha}_n
    + \vec\beta_{-n}\cdot\vec\beta_n+\vec{\tilde\beta}_{-n}\cdot\vec{\tilde\beta}_n \right]~
=~2\left(L_0+\tilde L_0\right).
\end{equation}
Physical states are constrained by the level matching  condition which is
\begin{equation}
  \Phi~=~\sum_{n\neq0} 
\left[ \vec\alpha_{-n}\cdot\vec\alpha_n - \vec{\tilde\alpha}_{-n}\cdot\vec{\tilde\alpha}_n
+    \vec\beta_{-n}\cdot\vec\beta_n -
\vec{\tilde\beta}_{-n}\cdot\vec{\tilde\beta}_n 
\right]~=~2\left(L_0-\tilde L_0\right)~ \sim 0 \, .
\label{lm}
\end{equation}
In both the classical and the quantum theory, the expression
(\ref{lm}) should vanish for physical states.  
Note that (\ref{lm}) is an exact expression that is independent of perturbation theory.  
It is obtained by plugging the oscillator expansion into the integral of the
constraint ${\cal S}=0$ over $\sigma$.  
In the quantum theory, level matching 
 is imposed as a physical state condition where (\ref{lm}) annihilates
physical states. 
One may check that $H_{1/4}, H_{1/8}$ and $H_{0}$ commute with
the level matching condition constraint $\Phi$ as it should. In the quantum theory which
we shall consider shortly, both $L_0$
and $\tilde L_0$ should be ambiguous up to a normal ordering constant.  However, because of 
the discrete symmetry which interchanges these operators, it would be reasonable that the constant
is the same for each operator and cancels in the difference $L_0-\tilde L_0$.

We therefore have the perturbative structure of the  squared space-time
Hamiltonian $p_{T}^2$
\begin{equation}\label{pttosecondorder}
p_T^2 = q_\phi^2 + \sqrt{\lambda} M^2 + \lambda^{1/4}\,H_{1/4}+ {\cal O}(\lambda^{1/8}),
\end{equation}
where, in terms of oscillators,
\bsp\label{Hinosc}
 H_{1/4} = &\vec p^2 + \vec q^2 + (\vec y^2 +\vec z^2)
(p_+^{(0)})^2 + \frac{(\vec z^2 -\vec y^2)}{2} M^2\\
&-\vec z^2 \sum_{n\neq 0} \vec\alpha_n \cdot \vec {\tilde\alpha}_{n}
+\vec y^2 \sum_{n\neq 0} \vec\beta_n \cdot \vec {\tilde\beta}_{n}
+ \frac{\vec z^2 + \vec y^2}{(p_+^{(0)})^2} \sum_{n}
L_{n}\tilde L_{n} 
 \, .
\end{split}
\end{equation}

Let us now quantize this system by promoting coordinates and modes to operators
and replacing $\{ . , ,\} \to -i [ . , ] $. We note the standard commutators
\begin{align}
[\alpha_{m}^{i},\alpha_{n}^{j}]&=m\,\delta_{m+n}\, \delta^{ij}\, , \qquad
[\tilde\alpha_{m}^{i},\tilde\alpha_{n}^{j}]=m\,\delta_{m+n}\, \delta^{ij}\, , \nn\\
[\beta_{m}^{i'},\beta_{n}^{j'}]&=m\,\delta_{m+n}\, \delta^{i'j'}\, , \qquad
[\tilde\beta_{m}^{i'},\tilde\beta_{n}^{j'}]=m\,\delta_{m+n}\, \delta^{i'j'}\, , \nn\\
[L_{m},\alpha_{n}^{i}]&=-n\,\alpha_{n+m}^{i}\, ,\qquad[L_{m},\beta_{n}^{i}]=-n\,\beta_{n_+m}^{i}\, ,\nn\\
[\tilde L_{m},\tilde\alpha_{n}^{i}]&=-n\tilde\alpha_{n+m}^{i}\, ,\qquad
[L_{m},\tilde\beta_{n}^{i}]=-n\,\tilde\beta_{n+m}^{i}\, .
\end{align}
It turns out that we can remove the last three terms in \eqn{Hinosc} through a unitary
transformation\footnote{Since $\hat V$ commutes with the operator $L_0-\tilde L_0$, it does not upset
the level matching condition.  It is easy to see that such a unitary transformation can be used to remove
any monomial in operators whose integer world-sheet momentum labels do not add to zero.  
The unitary transformation does not remove the term proportional to $L_0\tilde L_0$.},
\begin{equation}
\label{unitaryTransf}
\tilde p_{T}^{2} := e^{i\hat V/\lambda^{1/4}}\,  p_{T}^{2}\, e^{-i\hat V/\lambda^{1/4}}
= p_{T}^{2} + i\,\lambda^{1/4}\, [\hat V, M^{2}] + {\cal O}(\lambda^{0})\, ,
\end{equation}
where $\tilde p_{T}^2$ and $p_{T}^2$ have identical spectrum. Choosing the Hermitian operator $\hat V$ to be
\begin{equation}
\label{hatV1}
\hat V= -\frac{\vec z^{2}}{4}\sum_{n\neq0} \frac{i}{n}\, \vec\alpha_{n}\, \vec{\tilde\alpha}_{n}
+\frac{\vec y^{2}}{4}\sum_{n\neq0} \frac{i}{n}\, \vec\beta_{n}\, \vec{\tilde\beta}_{n}
+ \frac{\vec z^{2}+\vec y^{2}}{4(p_+^{(0)})^{2}}\sum_{n\neq0} \frac{i}{n}\, L_{n}\, \tilde L_{n}\, ,
\end{equation}
one can rotate away all non-diagonal terms at order $\lambda^{1/4}$, as\footnote{The contribution of the unitary transformation at  order $\lambda^0$ is  evaluated in appendix \ref{app:unit}}
\begin{equation}
\label{vm}
i\, [\hat V, M^{2}] = \vec z^2 \sum_{n\neq 0} \vec\alpha_n \cdot \vec {\tilde\alpha}_{n}
-\vec y^2 \sum_{n\neq 0} \vec\beta_n \cdot \vec {\tilde\beta}_{n}
- \frac{\vec z^2 + \vec y^2}{(p_+^{(0)})^{2}} \sum_{n\neq 0}
L_{n}\tilde L_{n} \, .
\end{equation}
We thus find 
\bsp \tilde p_{T}^{2} = &q_\phi^2 + \sqrt{\lambda}\, M^2\\ &+
\lambda^{1/4}\,\left(\vec p^2 + \vec q^2 + (\vec z^2+\vec y^2)
(p_+^{(0)})^{2} +\frac{(\vec z^2-\vec y^2)}{2} M^{2}+\frac{\vec z^2 +
\vec y^2}{(p_+^{(0)})^{2}}L_0\tilde L_0\right)\, + {\cal O}
(\lambda^{1/8}).
\end{split}
\end{equation}
The operator $L_0\tilde L_0$ can be  written as  $L_0\tilde L_0=\frac{1}{16}(M^4-\Phi^2)$, where  $\Phi$ is the level matching constraint. 
For physical states, obeying $\Phi|\text{phys}\rangle = 0$, we finally have 
\bsp
\tilde p_{T}^{2} = &q_\phi^2 + \sqrt{\lambda}\, M^2\\ &+ \lambda^{1/4}\,\left(\vec p^2 + \vec q^2 + \vec z^2\left( p_+^{(0)}+\frac{M^2}{4p_+^{(0)}}\right)^2
+ \vec y^2\left( p_+^{(0)}-\frac{M^2}{4p_+^{(0)}}\right)^2\right)\, 
+ {\cal O} (\lambda^{1/8})\, .
\end{split}
\end{equation}
This renders the Hamiltonian diagonal to this order and the spectrum
can be written down. We are interested in states where
$q_\phi$ is of order one. Therefore we must have that
\be
q_\phi = p_- + 2p_+ = {\cal O}(1),~~~
\text{and therefore}~~~p_+^{(0)} = \frac{M}{2},
\ee
where we are considering $M$ to be an eigenvalue. With this
restriction, we find
\be
\label{semifinal}
\tilde p_{T}^{2} = q_\phi^2 + \sqrt{\lambda}\, M^2 +
\lambda^{1/4}\,\left(\vec p^2 + M^2 \vec z^2 + \vec q^2 \right)
+ {\cal O} (\lambda^{1/8})\, ,
\ee
and identifying $(MR)^{2}\to\lambda^{1/2}\, M^{2}$, we find that at
${\cal O}(\l^{1/4})$ we have recovered precisely the particle energy
(\ref{enclassical}).

We work out the subleading terms $\lambda^{1/8}\, H_{1/8}$
and $\lambda^{0}\, H_{0}$ for ${\tilde p}^{2}_{T}$ of \eqn{semifinal} 
in the appendix. There we also show that
the $\lambda^{1/8}\, H_{1/8}$ term does not contribute to the spectrum
of $\tilde p_{T}$ down to order $\lambda^{0}$. Unfortunately the in
principle straightforward computation of correction to the spectrum 
\eqn{semifinal} at first order perturbation theory 
$\langle \text{phys} | H_{0} |\text{phys}\rangle$, in the sense
of  \eqn{sf2}, will depend on a large
number of so far unfixed normal ordering constants.

\section{Concluding remarks}

In this paper we have outlined an approach to the quantization of the
superstring in $AdS_5\times S^5$ in the flat-space limit. The first
step in taking the program further is to add the fermionic degrees of
freedom. In principle this should be a straightforward application of
the strategy employed here for the bosonic case. The reproduction of
the superparticle spectrum along the lines of section
\ref{sec:particle} should inform the correct scaling of the fermionic
fields, while the action itself is available for example from
\cite{Frolov:2006cc}. Another direction is to push the calculations
performed here to higher orders; this will likely require knowledge of
the fermionic terms in the higher-order Hamiltonian. The major
stumbling block in going to higher orders is the proliferation of
normal ordering constants. It would be nice to have a method of
determining these, perhaps through comparison to known results for
protected quantities or by matching against further numerical prediction
for higher excited states from the TBA and Y-system approach. 
As mentioned in the introduction, the higher
order Hamiltonian for which the bosonic contribution is provided in
appendix \ref{app:0}, will determine the ${\cal O}(\l^{-1/4})$
contribution to the energy $E$, once fermions have been added and
normal ordering constants have been determined. Given the unitary
transformation which has diagonalized the Hamiltonian at the previous
order, only first order perturbation theory is needed to extract the
answer. For bosonic external states, the contribution of fermionic
terms is relegated to those bosonic terms they produce through
self-contraction - i.e. normal ordering constant type terms. There may
be other methods of determining these, perhaps through closure of the
$PSU(2,2|4)$ quantum algebra.

Another puzzle is the interpretation of the zero-mode excitations. We
have argued here that we should take the zero-modes in their
ground-state in order to describe the length-four Konishi multiplet state. The
result for the energy $E$ at ${\cal O}(\l^0)$ is determined by the
zero-mode Hamiltonian. By exciting the zero-modes above the ground
state, one obtains non-zero results at ${\cal O}(\l^0)$. The dual
gauge theory interpretation of these states is still wanting. 
Given our interpretation, we may compare our results 
 at ${\cal O}(\l^0)$ to those of \cite{Roiban:2009aa}
where it is argued that the ${\cal O}(\l^0)$ term is $\D_0-4$, where
$\D_0$ is the bare dimension of the gauge theory operator. For the
length-four Konishi multiplet state one has $\D_0=4$, and so the absence of a
term at ${\cal O}(\l^0)$ argued in the introduction of this paper is
consistent with the results of \cite{Roiban:2009aa}.  It
may be that zero-mode excitations should be understood as
the string-duals of the longer members of the Konishi multiplet, for
which $\D_0-4 \geq 0$. It would be interesting to determine whether
this is indeed the case. An obvious problem with this interpretation is the 
unboundedness of the number of zero-mode oscillator excitations.

\section*{Acknowledgements}

We thank Gleb Arutyunov, Valentina Forini, 
Sergey Frolov, Elisabeth Kant, Volodya Kazakov, Per Sundin and Arkady Tseytlin for important discussions.

\appendix

\section{Higher order terms}

Considering  the expansion of $p_T^2$ out to the $\lambda^{0}$ order, it results
\begin{equation}
p_T^2 = q_\phi^2 + \sqrt{\lambda} M^2 + \lambda^{1/4}\,H_{1/4}+\lambda^{1/8}\,H_{1/8}+\lambda^0 H_0+\ldots
\end{equation}
where  $H_{1/8}$ and $H_{0}$ can be  computed  following the same  strategy described in section 3.  Applying the unitary transformation (\ref{unitaryTransf}) that diagonalizes  the  $\lambda^{1/4}$ term,  we  obtain 
\begin{align}\nonumber
\tilde p_{T}^{2}=& q_\phi^2 + \sqrt{\lambda}\, M^2 +
\lambda^{1/4}\,\left(\vec p^2 + M^2 \vec z^2 + \vec q^2 \right)+\lambda^{1/8}\,H_{1/8}\\ &+\lambda^0\left( H_0 + i[\hat V, H_{1/4}] -\frac{1}{2}[\hat V, [\hat V,M^2]]\right)+\ldots
\end{align}
In the  remainder of  this  appendix  we spell out  explicitly   $H_{1/8}$ and $H_{0}$   and the  $\lambda^0$ contribution of the  unitary transformation,  i.e.    $i[\hat V, H_{1/4}] -\frac{1}{2}[\hat V, [\hat V,M^2]]$.  

\subsection{$\lambda^{1/8}\, H_{1/8}$ term}
\label{app:18}

The  $H_{1/8}$ operator is given by 
\begin{align}
\label{cH18}
H_{1/8}
&=\int_0^{2\pi}\frac{d\sigma}{2\pi}\Bigg [
\frac{\vec{z}^2+\vec{y}^2}{4  (p_+^{(0)})^2}\, \Bigl [ \cM^{2}(\, \vec p\cdot
    \vec{\tilde{P}} +\vec q\cdot \vec{\tilde{Q}}\, ) - 2\, \cC\, (
    \vec{p}\cdot\vec{\tilde Z}' + \vec{q}\cdot\vec{\tilde Y}' )\,
    \Bigr ]
 +\vec{z}\cdot\vectt{Z}\, (\cM_\beta^2
  +2{\vectt{Z}'}^{2}) \nn\\ &-
  \vec{y}\cdot\vectt{Y}\, (\cM_\alpha^2 +2{\vectt{Y}'}^{2})
  +\frac{1}{8 (p_+^{(0)})^2}\, (\vec{y}\cdot\vectt{Y}+\vec{z}\cdot\vectt{Z})\,
  \Bigl [ \left (\cM^{2} \right)^{2 } 
    -4\cC^2\Bigr ]\Bigg ].
\end{align}
Every term in $H_{1/8}$  has  an odd number of oscillators. It therefore maps a state with an even number of oscillators onto a state with an odd number of oscillators.  For this
reason, the operator $H_{1/8}$ has vanishing matrix elements between all of the states of interest, that are two-oscillator
states of the  form   $\beta^i_{-1}\tilde\beta^j_{-1}|0\rangle$.   Therefore,    the   first order perturbation theory  correction  due to $H_{1/8}$,  which is of  order $\lambda^{\frac{1}{8}}$, vanishes.

The leading contribution due to $H_{1/8}$ is therefore in second order perturbation theory. Given that the energy denominators in second order perturbation theory  are always of order $\lambda^{\frac{1}{2}}$,\footnote{They are of  order $\lambda^{\frac{1}{2}}$ because the difference in the level number  between the states of interest
and the states that are created from them by operating with $H_{1/8}$, is non-zero.   In fact, it can be shown  that any term with an odd number of oscillators which respects level matching, when operating on $\beta_{-1}\tilde\beta_{-1}|0\rangle$
creates states whose level numbers differ by at least one from the level number of $\beta_{-1}\tilde\beta_{-1}|0\rangle$.} it results 
that this contribution is  of order $\left((\lambda)^{\frac{1}{8}}\right)^2\cdot \frac{1}{\lambda^{\frac{1}{2}}}\sim \lambda^{-\frac{1}{4}}$,  which is suppressed compared with the $\lambda^0$ terms
that are due to  first  order  perturbation  theory   of   $H_{0}$ and   the  unitary  transformation term.

\subsection{$\lambda^{0}\, H_{0}$ term}
\label{app:0}

Let us now work out the $H_{0}$ contributions. For convenience we split them into
four parts according to the order of oscillatory modes, i.e.
\begin{equation}
H_{0}=\int_0^{2\pi}\frac{d\sigma}{2\pi}\, \cH_{0} =
\int_0^{2\pi}\frac{d\sigma}{2\pi}\, \left( \cH_{0,0}+\cH_{0,2}+\cH_{0,4}+\cH_{0,6} \right)=
H_{0,0} + H_{0,2} + H_{0,4}+ H_{0,6}\, .
\end{equation}
Order zero in oscillatory modes:      
\begin{equation}
\cH_{0,0}=H_{0,0}=\frac{1}{2} \Bigl [\, (\vec{q}\cdot\vec{q}+4\, {p_+^{(2)}} {p_+^{(0)}}) \, \vec{z}\cdot\vec{z}
+(4\, {p_+^{(2)}} {p_+^{(0)}} - \vec{p}\cdot\vec{p}) \, \vec{y}\cdot\vec{y}\,\Bigr ] .
\end{equation}
Order two in oscillatory modes:
\begin{align}\label{H02}
\cH_{0,2}=& \frac{\vec{z}^{2}+\vec{y}^{2}}{4(p_+^{(0)})^2 }\, \Bigl [ 
(\vec{p}\cdot \vectt{P}+\vec{q}\cdot\vectt{Q})^{2} -
(\vec{p}\cdot\vectt{Z}'+\vec{q}\cdot\vectt{Y}')^{2}\, \Bigr ] +
\frac{  (\vec{z}^{2}+\vec{y}^{2})\, (\vec{p}^{2}
+\vec{q}^{2})}{8 (p_+^{(0)})^2}\,{\cal M}^2 \nn\\
& + (p_+^{(0)})^2\, (\vectt{Y}^{2}+\vectt{Z}^{2}) + 2\,
(\vec{q}\cdot\vectt{Q}\, \vec{z}\cdot\vectt{Z} -\vec{p}\cdot\vectt{P}\, \vec{y}\cdot\vectt{Y})
-\frac{1}{2}\, \vec{y}^{2}\, \vec{z}^{2}\, (\,{\vectt{Y}'}^{2}+{\vectt{Z}'}^{2}\, ) \nn \\
& \frac{(\vec{z}^{2})^{2}}{8}\, \Bigl [ -\frac{1}{2}\, \vectt{P}^{2}+\vectt{Q}^{2}+
{\vectt{Y}'}^{2}+\frac{9}{2}\, {\vectt{Z}'}^{2}\, \Bigr ]
+
\frac{(\vec{y}^{2})^{2}}{8}\, \Bigl [ -\frac{1}{2}\, \vectt{Q}^{2}+\vectt{P}^{2}+
{\vectt{Z}'}^{2}+\frac{9}{2}\, {\vectt{Y}'}^{2}\, \Bigr ] .
\end{align}
Order four in oscillatory modes:
\begin{align}\label{H04}
\cH_{0,4}=&
\frac{1}{2 (p_+^{(0)})^2 }\, \Bigl [ \cM^{2}\, (\vec{p}\cdot\vectt{P}+\vec{q}\cdot\vectt{Q})
-2\,\cC\, (\vec{p}\cdot\vectt{Z}'+\vec{q}\cdot\vectt{Y}')\,\Bigr ]\,(\vec{y}\cdot\vectt{Y}
+\vec{z}\cdot\vectt{Z}) \nn\\
& +\frac{1}{2}\, \Bigl[\, \vectt{Z}^{2}\, (\cM_{\beta}^{2}+2{\vectt{Z}'}^{2})
- \vectt{Y}^{2}\, (\cM_{\alpha}^{2}+2{\vectt{Y}'}^{2})\, \Bigr ]
-(\vec{z}^{2}+\vec{y}^{2})\,\frac{p_+^{(2)}}{8 (p_+^{(0)})^3}\, \Bigl [ (\cM^{2})^{2}-4\cC^{2}\, \Bigr ]
\nn\\
&+\frac{(\vec{z}^{2})^{2}}{4(p_+^{(0)})^2}\, \Bigl[ \frac{1}{4}\, \cM^{2}\, (\cM_{\beta}^{2}
+2\,{\vectt{Z}'}^{2})-\, \cC^{2}\Bigr ]
-\frac{(\vec{y}^{2})^{2}}{4(p_+^{(0)})^2}\, \Bigl[ \frac{1}{4}\, \cM^{2}\, (\cM_{\alpha}^{2}
+2\,{\vectt{Y}'}^{2}) -\cC^{2}\Bigr ]
\nn\\
& + \vec{z}^{2}\vec{y}^{2}\, \frac{\cM^{2}}{16 (p_+^{(0)})^2}\, (\cM_{\beta}^2+{2\vectt{Z}'}^{2}
-\cM_{\alpha}^{2} -2\,{\vectt{Y}'}^{2}) .
\end{align} 
Order six in oscillatory modes:
\begin{align}
\cH_{0,6}=&\frac{1} {128 (p_+^{(0)})^4}
(({\cM^2})^{2}-4 {\cC}^{2}) \Bigl [\, 8 (p_+^{(0)})^2\, (\vectt{Y}\cdot\vectt{Y} +\vectt{Z}\cdot\vectt{Z}) 
+{{\cM^2}} \, ( \vec{y}^2+ \vec{z}^2)^{2}\,\Bigr ] .
\end{align}
\subsubsection{Expressions in terms of oscillators}
The parts  of  $H_{0}$   can  be expressed in terms of  oscillators. 

Order two in oscillators:
\begin{equation}\label{H02osc}
\begin{split}
H_{0,2} &= \int_0^{2\pi}\frac{d\sigma}{2\pi} \cH_{0,2}=\\
&\frac{\vec{z}^{2}+\vec{y}^{2}}{2(p_+^{(0)})^{2}}\,\sum_{n\neq 0} \Bigl(
\vec p \cdot \alpha_n \, \vec p \cdot \tilde \alpha_n
+ \vec q \cdot \beta_n \, \vec q \cdot \tilde \beta_n
+\vec q \cdot \beta_n \, \vec p \cdot \tilde \alpha_n
+\vec p \cdot \alpha_n \, \vec q \cdot \tilde \beta_n \Bigr)\\
&+ \frac{1}{8 (p_+^{(0)})^2}\, (\vec{z}^{2}+\vec{y}^{2})\, (\vec{p}^{2}+\vec{q}^{2})\,M^2\\
&+ \frac{1}{2}\, (p_+^{(0)})^{2}\,\sum_{n\neq 0} \frac{1}{n^2} 
\Bigl( \alpha_n \cdot \alpha_{-n} + \tilde \alpha_n \cdot
  \tilde \alpha_{-n} - 2 \alpha_n \cdot \tilde \alpha_n
+\beta_n \cdot \beta_{-n} + \tilde \beta_n \cdot
  \tilde \beta_{-n} - 2 \beta_n \cdot \tilde \beta_n \Bigr)\\
&+i \sum_{n\neq 0} \frac{1}{n} \Bigl(-\vec q \cdot \beta_n\, \vec z \cdot \alpha_{-n}
+ \vec q \cdot \beta_n \, \vec z \cdot \tilde \alpha_n
+ \vec q \cdot  \tilde\beta_n \, \vec z \cdot \alpha_n
-\vec q \cdot\tilde \beta_n\, \vec z \cdot \tilde\alpha_{-n}\\
&\qquad\qquad\qquad+ \vec p \cdot \alpha_n\, \vec y \cdot \beta_{-n}
- \vec p \cdot \alpha_n \, \vec y \cdot \tilde \beta_n
- \vec p \cdot  \tilde\alpha_n \, \vec y \cdot \beta_n
+\vec p \cdot\tilde \alpha_n\, \vec y \cdot \tilde\beta_{-n} \Bigr)\\
&+\frac{ ({\vec z}^2)^2 + ({\vec y}^2)^2}{8} M^2 
-\frac{3}{8}  ({\vec z}^2)^2 \sum_{n\neq 0} \alpha_n \cdot \tilde \alpha_n
-\frac{3}{8}  ({\vec y}^2)^2 \sum_{n\neq 0} \beta_n \cdot \tilde \beta_n\\ 
&+ \frac{1}{4} {\vec z}^2 \left(\frac{1}{2} {\vec z}^2 - {\vec y}^2
\right)\sum_{n\neq 0} \left( \alpha_n \cdot \alpha_{-n} + \tilde \alpha_n \cdot
  \tilde \alpha_{-n} - 2 \alpha_n \cdot \tilde \alpha_n \right)\\
&+ \frac{1}{4} {\vec y}^2 \left(\frac{1}{2} {\vec y}^2 - {\vec z}^2
\right)\sum_{n\neq 0} \left( \beta_n \cdot \beta_{-n} + \tilde \beta_n \cdot
  \tilde \beta_{-n} - 2 \beta_n \cdot \tilde \beta_n \right).
\end{split}
\end{equation}
Order six in oscillators:
\begin{equation}
\begin{split}
H_{0,6} &=\int_0^{2\pi}\frac{d\sigma}{2\pi} \cH_{0,6}=\\ 
&-\frac{1}{2(p_+^{(0)})^2} \sum_{\substack{n+m+p+q=0\\p,q\neq 0}}
 \frac{1}{pq} \left( \alpha_p \cdot \alpha_q + \tilde \alpha_{-p}
   \cdot \tilde \alpha_{-q} - 2 \alpha_p \cdot \tilde \alpha_{-q}
 \right) L_n \tilde L_{-m}\\
&-\frac{1}{2(p_+^{(0)})^2} \sum_{\substack{n+m+p+q=0\\p,q\neq 0}}
 \frac{1}{pq} \left( \beta_p \cdot \beta_q + \tilde \beta_{-p}
   \cdot \tilde \beta_{-q} - 2 \beta_p \cdot \tilde \beta_{-q}
 \right) L_n \tilde L_{-m}\\
&+\frac{({\vec z}^2 +{\vec y}^2)^2}{4(p_+^{(0)})^4} \sum_{n+m+p=0} 
L_n \tilde L_{-m} \left( L_p + \tilde L_{-p} \right).
\end{split}
\end{equation}
Order four in oscillators:
\begin{equation}\label{H04osc}
\begin{split}
H_{0,4} &=\int_0^{2\pi}\frac{d\sigma}{2\pi} \cH_{0,2}=\\ 
&\frac{1}{(p_+^{(0)})^2}\sum_{\substack{n+m+p=0\\m,p\neq 0}} \frac{i}{p}
\Bigl[ L_n ( \vec p \cdot \tilde \alpha_{-m} +
\vec q \cdot \tilde \beta_{-m} )
+\tilde L_{-n} (\vec p \cdot  \alpha_{m} +
\vec q \cdot \beta_{m} ) \Bigr]\\ &\qquad\qquad\qquad\qquad\qquad \times\left( \vec z \cdot \alpha_p - \vec z \cdot \tilde \alpha_{-p}
+ \vec y \cdot \beta_p - \vec y \cdot \tilde \beta_{-p}\right)\\
&-\frac{1}{2} \sum_{\substack{n+m+p=0\\m,p\neq 0}} \frac{1}{mp}
\left( \alpha_p \cdot \alpha_m + \tilde \alpha_{-p} \cdot \tilde \alpha_{-m}
  -2  \alpha_p \cdot \tilde \alpha_{-m} \right) \left( L_n + \tilde
  L_{-n} \right)\\
&+\frac{1}{2} \sum_{\substack{n+m+p+q=0\\n,m,p,q\neq 0}} \frac{1}{mp}
\left( \alpha_p \cdot \alpha_m + \tilde \alpha_{-p} \cdot \tilde \alpha_{-m}
  -2  \alpha_p \cdot \tilde \alpha_{-m} \right) \, \alpha_n \cdot
\tilde \alpha_{-q}\\
&+\frac{1}{2} \sum_{\substack{n+m+p=0\\m,p\neq 0}} \frac{1}{mp}
\left( \beta_p \cdot \beta_m + \tilde \beta_{-p} \cdot \tilde \beta_{-m}
  -2  \beta_p \cdot \tilde \beta_{-m} \right) \left( L_n + \tilde
  L_{-n} \right)\\
&-\frac{1}{2} \sum_{\substack{n+m+p+q=0\\n,m,p,q\neq 0}} \frac{1}{mp}
\left( \beta_p \cdot \beta_m + \tilde \beta_{-p} \cdot \tilde \beta_{-m}
  -2  \beta_p \cdot \tilde \beta_{-m} \right) \, \beta_n \cdot
\tilde \beta_{-q}\\
&-2p_+^{(2)}\frac{{\vec z}^2 + {\vec y}^2}{(p_+^{(0)})^3} 
\sum_{n} L_n \tilde L_{n}\\
&+\frac{({\vec z}^2)^2-({\vec y}^2)^2}{ (p_+^{(0)})^2} \sum_{n} L_n \tilde L_{n}
-\frac{({\vec z}^2)^2}{4 (p_+^{(0)})^2}\sum_{\substack{n+m+p=0\\m,p\neq 0}} \left(
L_n + \tilde L_{-n} \right) \alpha_{m} \cdot \tilde \alpha_{-p} \\
&+\frac{({\vec y}^2)^2}{4 (p_+^{(0)})^2 }\sum_{\substack{n+m+p=0\\m,p\neq 0}} \left(
L_n + \tilde L_{-n} \right) \beta_{m} \cdot \tilde \beta_{-p} \\
&- \frac{{\vec z}^2 {\vec y}^2}{4 (p_+^{(0)})^2} \sum_{\substack{n+m+p=0\\m,p\neq 0}} \left(
L_n + \tilde L_{-n} \right)\left( \alpha_{m} \cdot \tilde \alpha_{-p}
- \beta_{m} \cdot \tilde \beta_{-p} \right),
\end{split}
\end{equation}

\subsection{$\mathcal{O}(\lambda^0)$  contribution of the unitary transformation}
\label{app:unit}

We now  compute the   $\lambda^0$ order contribution of  the unitary transformation (\ref{unitaryTransf}) given by the term
 \begin{equation}
 i[\hat V, H_{1/4}] -\frac{1}{2}[\hat V, [\hat V,M^2]].
\end{equation}
Note that as there are no ordering ambiguities in the $M^{2}$ and $H_{1/4}$ operators,
this $\mathcal{O}(\lambda^{0})$ contribution from the unitary transformation does not
suffer from ordering ambiguities.

The relevant commutators  are
\begin{equation}
\begin{split}
&[{\vec z}^2,{\vec p}^2] = 4i \vec z \cdot \vec p + 8,\quad
[{\vec y}^2,{\vec q}^2] = 4i \vec y \cdot \vec q + 8,\\
&[\alpha_n \cdot \tilde \alpha_n, :L_0:+:\tilde L_0:] =  2\, n\, \alpha_n \cdot
\tilde \alpha_n,\quad
[\beta_n \cdot \tilde \beta_n, :L_0:+:\tilde L_0:] = 2\, n\, \beta_n \cdot
\tilde \beta_n,\\
&[\alpha_n \cdot \tilde \alpha_n, L_m \tilde L_m] 
= n \,\alpha_{m+n} \cdot \tilde \alpha_n \tilde L_m +
n \,L_m\,\tilde \alpha_{m+n} \cdot \alpha_n, \\
&[\beta_n \cdot \tilde \beta_n, L_m \tilde L_m] 
= n \,\beta_{m+n} \cdot \tilde \beta_n \tilde L_m +
n \,L_m\,\tilde \beta_{m+n} \cdot \beta_n, \\
& [L_m \tilde L_m , :L_0:+:\tilde L_0:] = 2m\,L_m\tilde L_m,\\
&\left[ \sum_{n\neq 0} \frac{i}{n} \alpha_n \cdot \tilde 
  \alpha_n , \sum_{m\neq 0}  \alpha_m \cdot \tilde 
  \alpha_m \right] = i \sum_{n\neq 0} \left( \alpha_{-n}\cdot
  \alpha_n + \tilde \alpha_n \cdot \tilde \alpha_{-n}\right),\\
&\left[ \sum_{n\neq 0} \frac{i}{n} \beta_n \cdot \tilde 
  \beta_n , \sum_{m\neq 0}  \beta_m \cdot \tilde 
  \beta_m \right] = i \sum_{n\neq 0} \left( \beta_{-n}\cdot
  \beta_n + \tilde \beta_n \cdot \tilde \beta_{-n}\right),\\
&[L_m \tilde L_m,  L_n \tilde L_n] = (m-n) \, \left( \tilde L_n \tilde L_m
L_{m+n} + L_m L_n \tilde L_{m+n} \right) + \frac{c_V}{12}(m^3-m)(L_mL_n+\tilde L_n \tilde L_m)\delta_{m+n} ,
\end{split}
\end{equation}
where we used 
\begin{align}
[L_m,  L_n ] &= (m-n) \,
L_{m+n}  + \frac{c_V}{12}(m^3-m)\delta_{m+n}\nn\\
[\tilde L_m,  \tilde L_n] &= (m-n) \, \tilde L_{m+n}  + \frac{c_V}{12}(m^3-m)\delta_{m+n}
\end{align}
We find that 
\begin{equation}
\begin{split}
 i[\hat V, H_{1/4}] -&\frac{1}{2}[\hat V, [\hat V,M^2]] =\\ &
(2 +i \vec z \cdot \vec p) \sum_{n\neq 0} \frac{1}{n} \alpha_n
\cdot \tilde \alpha_n 
- (2 + i \vec y \cdot \vec q) \sum_{n\neq 0} \frac{1}{n} \beta_n
\cdot \tilde \beta_n \\
&- \bigl( i (  \vec z \cdot \vec p + \vec y \cdot \vec q ) +
  4\bigr) \,\frac{1}{(p_+^{(0)})^{2}} \sum_{n\neq 0} \frac{1}{n} L_n \tilde L_n\\
&+  \frac{{\vec z}^2({\vec z}^2-{\vec y}^2)}{2} \sum_{n\neq 0} \alpha_n\cdot \tilde \alpha_n
 -  \frac{{\vec y}^2({\vec z}^2-{\vec y}^2)}{2} \sum_{n\neq 0} \beta_n\cdot \tilde \beta_n -  \frac{  ({\vec z}^2)^2 - ({\vec y}^2)^2}{2(p_+^{(0)})^{2}} \sum_{n\neq 0}  L_n\tilde L_n\\
& - \frac{({\vec z}^2)^2}{8} \sum_{n\neq 0} \left( \alpha_{-n}\cdot
  \alpha_n + \tilde \alpha_n \cdot \tilde \alpha_{-n}\right)
- \frac{({\vec y}^2)^2}{8} \sum_{n\neq 0} \left( \beta_{-n}\cdot
  \beta_n + \tilde \beta_n \cdot \tilde \beta_{-n}\right)\\
&+  \frac{{\vec z}^2 ( {\vec z}^2 + {\vec y}^2)}{8(p_+^{(0)})^{2}} \sum_{n\neq 0,\,
  m\neq 0} \left(1-\frac{n}{m}\right)\left( \alpha_{m+n} \cdot \tilde \alpha_n \, \tilde L_m +
  L_m \tilde \alpha_{m+n} \cdot \alpha_n \right)\\
&-  \frac{{\vec y}^2 ( {\vec z}^2 + {\vec y}^2)}{8(p_+^{(0)})^{2}} \sum_{n\neq 0,\,
  m\neq 0} \left(1-\frac{n}{m}\right)\left( \beta_{m+n} \cdot \tilde \beta_n \, \tilde L_m +
  L_m \tilde \beta_{m+n} \cdot \beta_n \right)\\
&-\frac{({\vec z}^2 + {\vec y}^2)^2}{8(p_+^{(0)})^4} \sum_{n\neq 0,\, m\neq 0}
 \left(1 - \frac{n}{m}\right) \left( \tilde L_n \tilde L_m
L_{m+n} + L_m L_n \tilde L_{m+n} \right)\\
&-\frac{c_V({\vec z}^2 + {\vec y}^2)^2}{96 (p_+^{(0)})^4} \sum_{n\neq 0}(n^2-1)(L_{n}L_{-n}+\tilde L_{-n} \tilde L_{n})\\
&+\frac{{\vec z}^2({\vec z}^2 + {\vec y}^2)}{4(p_+^{(0)})^2}\sum_{n\neq 0} ((:L_0:+c_L)\alpha_n\cdot \tilde \alpha_n+\alpha_n\cdot \tilde \alpha_n(:\tilde L_0:+\tilde c_L))\\
&-\frac{{\vec y}^2({\vec z}^2 + {\vec y}^2)}{4(p_+^{(0)})^2}\sum_{n\neq 0} ((:L_0:+c_L)\beta_n\cdot \tilde \beta_n+\beta_n\cdot \tilde \beta_n(:\tilde L_0:+\tilde c_L))\\
&-\frac{({\vec z}^2 + {\vec y}^2)^2}{4(p_+^{(0)})^4}\sum_{n\neq 0} ((:L_0:+ c_L) L_n\cdot \tilde L_n+L_n\cdot \tilde L_n(:\tilde L_0:+\tilde c_L))
\end{split}
\end{equation}
%

\bibliographystyle{nb}
\bibliography{botany}

\end{document}